\documentclass{article}
\usepackage{spconf,amsmath,graphicx}
\usepackage{bm, amssymb, booktabs, multirow,cite, url, color}

\title{Dynamic Acoustic Compensation and Adaptive Focal Training for Personalized Speech Enhancement}

\name{Xiaofeng Ge$^{1}$, Jiangyu Han$^{1}$, Haixin Guan$^{2}$, Yanhua Long$^{1}$\sthanks{Yanhua Long is the corresponding author. The work is supported by the National Natural Science Foundation of China (Grant No.62071302 and No.61701306).}}
 
\address{
  $^1$Key Innovation Group of Digital Humanities Resource and Research, \\
     Shanghai Normal University, Shanghai, China\\
  $^2$Unisound AI Technology Co., Ltd., Beijing, China}

\begin{document}
\ninept
\maketitle

\begin{abstract}


Recently, more and more personalized speech enhancement systems (PSE)
with excellent performance have been proposed. However, two critical issues
still limit the performance and generalization ability of the model:
1) Acoustic environment mismatch between the test noisy speech and 
target speaker enrollment speech; 2) Hard sample mining and learning.
In this paper, dynamic acoustic compensation (DAC) is proposed to alleviate the
environment mismatch, by intercepting the noise or environmental
acoustic segments from noisy speech and mixing it with the clean 
enrollment speech. To well exploit the hard samples in training data,
we propose an adaptive focal training (AFT) strategy by assigning
adaptive loss weights to hard and non-hard samples during training.
A time-frequency multi-loss training is further introduced to improve and generalize our previous
work sDPCCN for PSE. The effectiveness of proposed
methods are examined on the DNS4 Challenge dataset. Results show that,
the DAC brings large improvements in terms of multiple
evaluation metrics, and AFT reduces the hard sample rate significantly and produces obvious MOS score improvement.

\end{abstract}
\begin{keywords}
personalized speech enhancement, acoustic environment mismatch, hard sample, dynamic acoustic compensation, adaptive focal training
\end{keywords}
\vspace{-0.2cm}
\section{Introduction}
\label{sec:intro}

Personalized speech enhancement (PSE) aims to extract and enhance the target
speaker's speech in a complicated multi-talker noisy or reverberant acoustic environment.
Recently, there has been increasing attention to PSE, and some competitions
such as the 4th Deep Noise Suppression (DNS4) Challenge \cite{dns}
has set up a specialized track for this task. According to DNS4 Challenge,
we can conclude that an excellent PSE system is generally required to enhance
the target speech with the help of target speaker’s enrollment speech
in three different situations: 1) Noisy with no interfering speaker;
2) Clean but with interfering speakers. 3) Noisy and with interfering speakers.

In recent years, deep learning based personalized speech enhancement
has achieved promising performance \cite{tea, model2, model3}.
However, there are still two problems need to be solved.
The first one is the acoustic environment mismatch between
the enrollment speech and the noisy speech to be enhanced.
Generally, the provided enrollment speech is clean and the speech
to be enhanced may contain serious noise, such mismatch makes the reference
voiceprint information misleading to some extent.
The second problem we found is the hard sample mining and learning.
According to objective evaluation metrics, samples with low scores
can be regarded as hard and the others as non-hard. Although hard samples
tend to have greater losses during training, the model is still biased
towards non-hard samples because hard samples are often sparse in
training set. That’s why many models can achieve excellent average
evaluation scores on the whole test set, but have very poor performance
on a few hard samples. In fact, such hard samples will seriously
affect the user experience of PSE system in real scenarios,
which makes it is meaningful and important to improve the system enhancement
ability on hard samples contrapuntally.

Although there are many previous works focused on acoustic mismatch
between training and test datasets \cite{mis1}, or even cross-domain speech
extraction tasks \cite{mis2}, the works on dealing with acoustic environment mismatch between
test noisy and speaker enrollment speech is limited. Authors in \cite{4} have proposed an iterative
refined adaption method, it uses the original speaker embedding to extract target speech first
and re-encodes the extracted speech to obtain a refined text-dependent speaker embedding.
Then, the refined embedding is combined with original one using weights to
form a new embedding to extract the target speech again. This method can lead to
good results, but it heavily increased the computational complexity and inference time.
The hard sample issue has attracted much attention in computer vision,
such as the re-weighting or novel loss function \cite{hs1, hs2, hs3}, etc.
However, it is very difficult to find works on hard sample problem in personalized
speech enhancement in the literature. Therefore, we think that it will be
interesting to explore methods for hard sample mining in PSE tasks.

In \cite{7}, our previous proposed sDPCCN was specially designed
for pure target speech extraction and achieved excellent performance and
robustness. In this study, we first generalize sDPCCN to
the personalized speech enhancement task, then three new improvements
are proposed to further enhance the sDPCCN for PSE: 1) TF-loss.
The original sDPCCN only uses the time-domain negative scale
invariant signal-to-noise ratio (SISNR) \cite{sisnr} loss, here we
combine the SISNR with a frequency-domain mean square error (MSE)
loss to leverage the information from different domains during model
training; 2) Dynamic acoustic compensation (DAC). The DAC is designed to dynamically
update the background acoustic characteristics of target
speaker enrollment speech with each input test noisy speech, by
simply intercepting the noise or environmental acoustic
segments from noisy test speech and mixing it with the clean
enrollment utterance; 3) Adaptive focal training (AFT).
The AFT is proposed to emphasize the importance of hard samples
in training data by using an adaptive focal loss during model
training, where a $sin(\cdot)$ transformation is used to weight
each batch normalized TF-loss. All our experiments are performed
on the DNS4 dataset, results show that all the proposed methods are
effective, especially the DAC and AFT. They make the model outperform 
sDPCCN baseline significantly in both signal distortion and
perceptually evaluation metrics.

\vspace{-0.2cm}
\section{sDPCCN}
\vspace{-0.1cm}
\label{sec:sDPCCN}

Our previous work DPCCN \cite{7}, a densely-connected pyramid complex convolutional network,
is a U-Net \cite{unet} style network that proposed for speech separation. And we extended
it to sDPCCN for target speech extraction (TSE), by simply integrating a small block of
enrollment speaker encoder to supervise the network towards extracting target speech. Extensive
experiments in \cite{7} showed that both DPCCN and sDPCCN achieved good performance and
robustness, surpassing some well-known systems such as Conv-TasNet \cite{convtas} and
TD-SpeakerBeam \cite{tsb} on speech separation and TSE task respectively. In these two networks, DenseNet \cite{densenet}
is used in encoder/decoder to alleviate the vanishing-gradient problem and encourage
the feature reuse. Temporal convolutional network (TCN) \cite{tcn} is used as the
bottleneck layer between encoder and decoder to capture long-range time information.
A novel pyramid pooling layer \cite{pyramid} introduced after the decoder is used to
leverage the global information. Due to the promising performance of sDPCCN in TSE task, 
we generalize it to the personalized speech enhancement.
More details of DPCCN and sDPCCN can be found in \cite{7}.


\vspace{-0.2cm}
\section{Proposed Methods}
\vspace{-0.1cm}
\label{sec:proposed_methods}

\begin{figure}[t]
  \centering
  \setlength{\abovecaptionskip}{0.cm}
  \includegraphics[width=\linewidth]{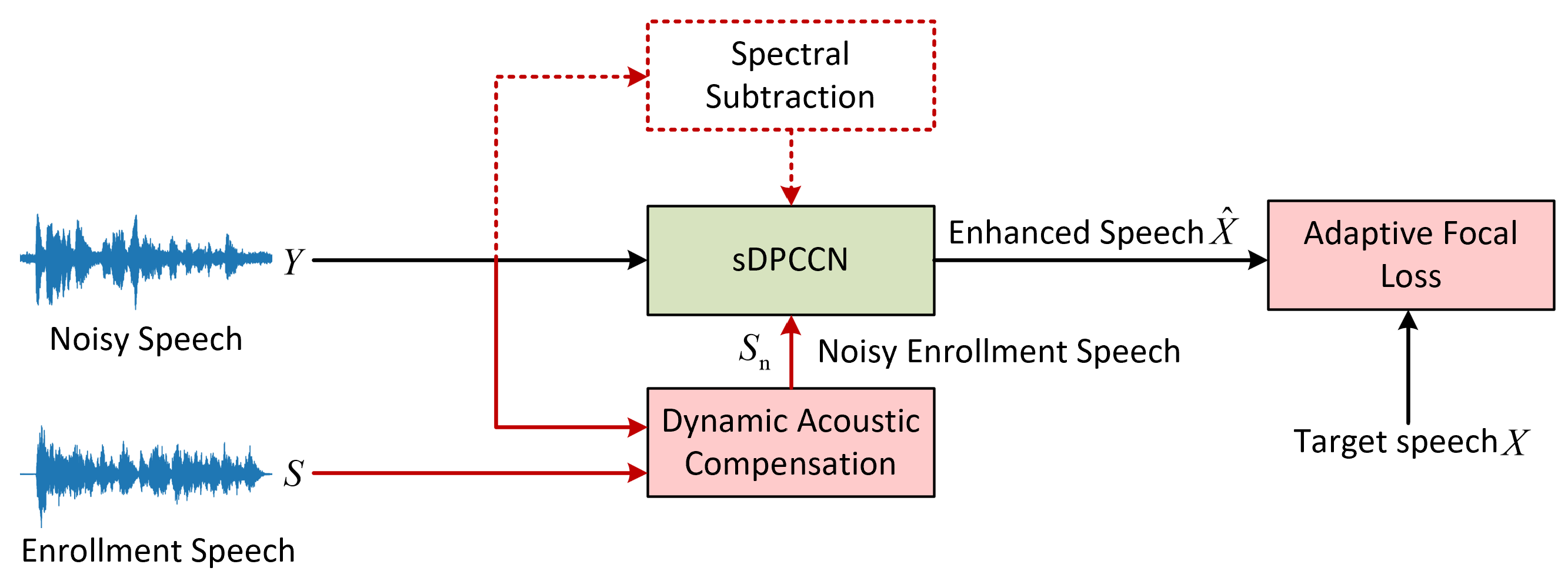}
  \caption{The framework of proposed algorithm, all red blocks and lines
  are our improvements. The dashed block is the algorithm that alleviate acoustic environment mismatch by spectral subtraction, which is used for comparison with DAC.}
  \label{fig:dpccn}
  \vspace{-0.2cm}
\end{figure}
\vspace{-0.2cm}

\subsection{Architecture}
\label{ssec:dpccn}
\vspace{-0.1cm}

The whole architecture of our proposed methods is shown in Fig.\ref{fig:dpccn}.
We generalize the sDPCCN \cite{7} to personalized speech enhancement with
several innovations, including the TF-loss, the dynamic acoustic compensation,
and the adaptive focal training. For performance comparison, the traditional
pre-processing methods: spectral subtraction \cite{6} and MMSE-LSA \cite{lsa}
are also investigated to deal with the acoustic environment mismatch issue.
Details of the new methods are presented as follows.

\vspace{-0.2cm}
\subsection{TF-loss}
\vspace{-0.1cm}
\label{ssec:tf}


The loss function used in original sDPCNN is the negative SISNR \cite{sisnr},
which is commonly used in speech separation and target speech extraction tasks. Since we generalize sDPCCN to personalized speech enhancement,
it is reasonable to combine the loss which is commonly used in speech enhancement
with the negative SISNR. In addition, many previous works \cite{cd1, cd2, cd3}
have proved that it is beneficial to leverage information from multiple domains
when computing the system training loss. Based on these two points,
we choose the most commonly used mean square error (MSE) loss $\mathcal{L}_{MSE}$ to provide
complementary information for speech enhancement. The combined training
loss $\mathcal{L}_{TF}$ is defined as:
\begin{equation}
\label{eq:ltf}
  {\mathcal{L}_{TF}} = \mathcal{L}_{-SISNR} + \mathcal{L}_{MSE}.
\end{equation}
where the negative SISNR $\mathcal{L}_{-SISNR}$ is computed in time domain
while $\mathcal{L}_{MSE}$ is computed in frequency domain (the output before iSTFT block in sDPCCN).

\vspace{-0.1cm}
\subsection{Dynamic Acoustic Compensation}
\vspace{-0.1cm}
\label{ssec:mix-remix}

\begin{figure}[t]
  \centering
  \includegraphics[width=\linewidth]{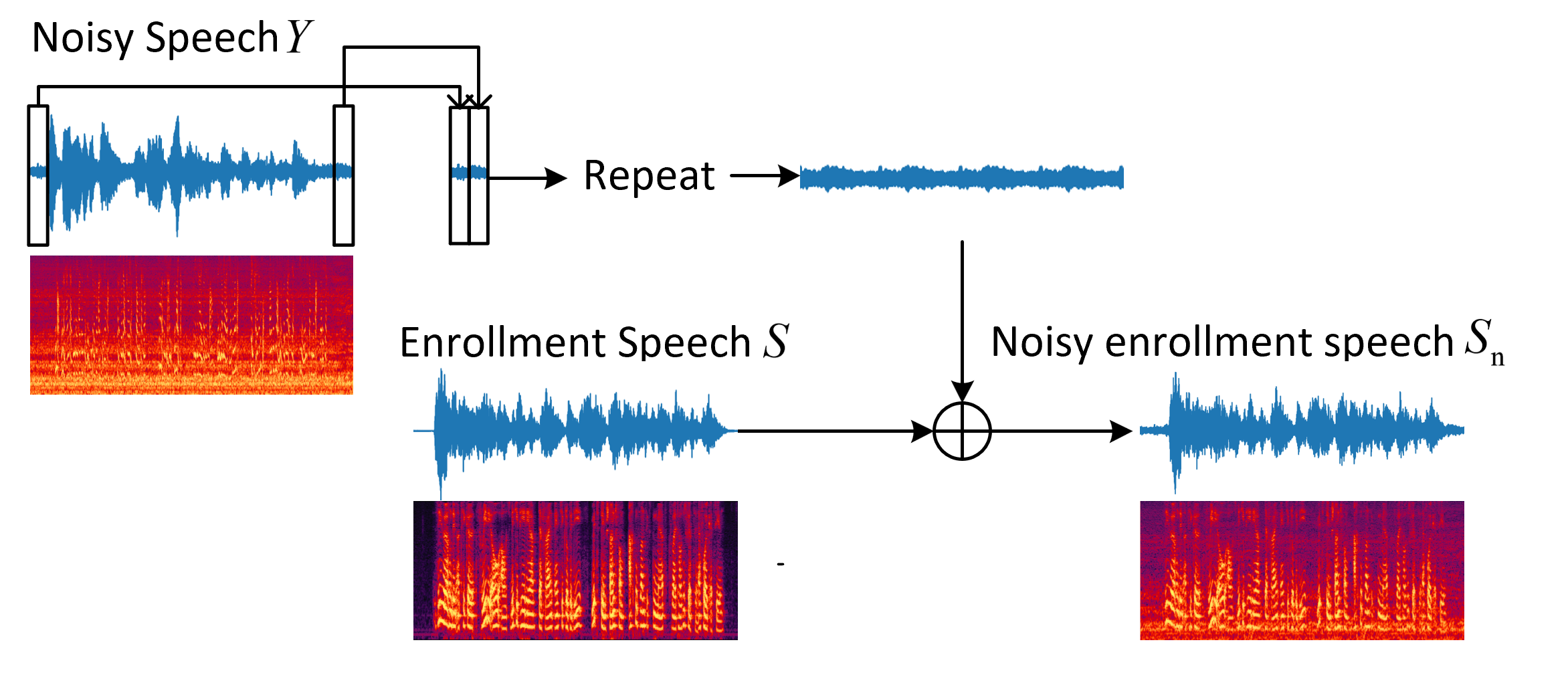}
  \caption{Illustration of dynamic acoustic compensation (DAC).}
  \label{fig:SA}
  \vspace{-0.5cm}
\end{figure}

As shown in Fig.\ref{fig:dpccn} and Fig.\ref{fig:SA}, given a random test noisy
speech $Y$, the acoustic environment mismatch between $Y$ and enrollment
speech $S$ varies with $Y$ dynamically during speech enhancement.
This is because it is almost impossible for two non-simultaneous
recorded speech to be in the same acoustic environment. The acoustic
mismatch between noisy $Y$ and enrollment speech $S$ is unfavorable but inevitable,
which will lead to a poor PSE performance.

To alleviate this problem, the dynamic acoustic compensation (DAC) is proposed
and illustrated in Fig.\ref{fig:SA}. Given $S$ and a random noisy input
speech waveform $Y$, we first simply intercept the first $J$ and the last $K$
frames signal of $Y$ as the background noise. Then we concatenate and repeat
them to form a signal with the same length as enrollment speech $S$. The final
noisy enrollment speech $S_{n}$ is obtained as,
\begin{equation}
\label{eq:tf}
  S_{n} = repeat(Y_{1},\ldots,Y_{J},Y_{T-K+1},\ldots,Y_{T}) + S
\end{equation}
where $T$ is the total frame length of $Y$.
Actually, our DAC is motivated by the traditional spectral subtraction (SS)
algorithm \cite{6}. Although it is very simple, it effectively compensates for the acoustic background characteristics of clean
enrollment speech in a dynamic manner, which makes the result $S_{n}$ contains
similar noise to $Y$ to avoid the heavy acoustic environment mismatch.
As different input noisy speech may be recorded under different background
environments, the advantage of DAC makes it very flexible for
real PSE scenarios. Acorrding to our experiments, we have found that our proposed DAC works well in most cases and there will not
be any negative effect during the whole speech enhancement process.



In addition, from the opposite perspective, we can remove the background
noise in $Y$ to eliminate the environment mismatch with $S$.
Such as, using the traditional spectral addition and MMSE-LSA \cite{lsa} as
pre-processing to remove noise before passing $Y$ to sDPCCN.
However, these methods may lead to unrecoverable speech distortion
and lower the performance.


\vspace{-0.2cm}
\subsection{Adaptive Focal Training}
\vspace{-0.1cm}


In general, the number of hard samples in training set is
sparse compared with the non-hard samples, so the model biases
towards the non-hard samples in order to reduce the overall
loss during training. Hard samples are often ignored because their less
contribution to the average performance of the whole test set.
However, when a PSE system encounters such a hard sample, its bad
performance will heavily affect the human auditory perception.
Therefore, to improve hard sample enhancement quality,
we propose a novel two-stage training strategy,
termed adaptive focal training (AFT), by introducing an adaptive
focal loss $\mathcal{L}_{AFT}$ for each batch as:
\begin{equation}
\label{eq:aft}
  {\mathcal{L}_{AFT}} = \sum_{i=1}^{B} \left(\mathcal{L}_{TF}^{i}*\sin\left(\frac{\pi}{2}*\left(\frac{\mathcal{L}_{TF}^{i}- \mu}{\sigma}\right)\right)\right)
\end{equation}
where $B$ is the batch size of training data, $\mathcal{L}_{TF}^{i}$ is the TF-loss
defined in Eq.(\ref{eq:tf}) of $i$-th sample in current batch, $\mu$ and $\sigma$
are the mean and standard deviation of  $\mathcal{L}_{TF}$ in each batch.

\verb"Two-stage training": Using $\sin(\cdot)$ transformation in
Eq.(\ref{eq:aft}) can assign a larger weight to $\mathcal{L}_{TF}$
of hard samples than to non-hard samples during training, because
hard samples often tend to have larger $\mathcal{L}_{TF}$ than non-hard
ones. However, if we use $\mathcal{L}_{AFT}$ directly during the whole
training, it will make the model pay too much attention to hard
samples and neglect the major non-hard samples. Therefore, the
two-stage training is proposed in AFT to obtain a more balanced model: 1)
\verb"Stage1": performing $\mathcal{L}_{TF}$  for training to make model
focus on major non-hard samples, the early-stop training
\cite{early} is used to control the epochs in this stage; 2) \verb"Stage2": 
continue to train the stopped model using $\mathcal{L}_{AFT}$ for more epochs
($<$20) until model converge. We think that the model in \verb"Stage1"
is good enough to learn the major non-hard samples, and in \verb"Stage2"
$\mathcal{L}_{AFT}$ can alleviate the model bias towards non-hard samples that
generated in the first stage and result in a more balanced PSE model.

\vspace{-0.3cm}
\section{Experimental Setup}
\label{sec:task}
\vspace{-0.2cm}

\subsection{Datasets}
\vspace{-0.1cm}

The clean speech, noise, and room impulse response (RIR) we used for
creating training data are all from DNS4 Challenge dataset \cite{dns}.
Specifically, the clean speech includes six languages
(English, French, German, Italian, Russian, and Spanish)
from 3,230 speakers in total. Noise data mainly comes from
Audioset \cite{data1}, Freesound\footnote{https://freesound.org/},
and DEMAND \cite{data2}. Since our purpose is to verify the effectiveness
of the proposed methods, instead of using the entire DNS4 Challenge dataset,
we only create a 160 hours training set, covering three PSE conditions: 1) noise: 100 hours with only
target speaker and background noise; 2) mix: 40 hours with target speaker and
one interfering speaker, and 3) nmix: 20 hours with target, one interfering speaker
and background noise. The noisy speech segments are all
simulated by dynamically mixing speech and noise (or interfering speech)
with a random SNR ranges from -5 to 20 dB, the RT60 of RIRs ranges
from 0.3s to 1.3s.


We also use part of DNS4 Challenge dataset to simulate a test set,
termed DNS-test, without any overlap with the 160 hours training data.
DNS-test consists of 800 samples, and can also be divided into the
same three PSE conditions in the same order as mentioned above:
t-noise (500 samples), t-mix (200 samples) and t-nmix (100 samples).
The duration of each sample is 10 seconds. The data proportion of the
three conditions contained in our both training set and DNS-test is
similar as it of official DNS4-Challenge test set.

%

\vspace{-0.2cm}
\subsection{Configuration}
\vspace{-0.1cm}

All training and test data are sampled at 8kHz. For STFT,
we use the square root of Hanning window with FFT size of 512 and
hop size of 128. Global mean-variance normalization is
applied to all input features. Batch size is set to 32.
We train all of our models with Adam optimizer \cite{adam}
and the initial learning rate is set to 0.001.
When the loss of the validation set does not decreases
for 3 epochs, the learning rate will be decayed by a
ratio of 0.5. To make experimental results comparable,
all the other configurations are the same as the original sDPCCN in \cite{7}.

\vspace{-0.2cm}
\subsection{Evaluation Metrics}
\vspace{-0.1cm}

SISNR, PESQ \cite{pesq} and DNSMOS \cite{dnsmos} are used to measure
the audio quality comprehensively. DNSMOS is estimated by a trained deep learning model provided by DNS4 Challenge to approximately represent the traditional MOS \cite{mos} score, and it consists of three scores that
measure speech quality (SIG), background noise quality (BAK), and overall
audio quality (OVR) respectively. In addition to these widely
used objective metrics, inspired by \cite{hsr},
we also use hard sample rate (HSR) to checkout the effectiveness
of adaptive focal training. For example, HSR0 (\%) means
the proportion of enhanced samples with SNR lower than 0dB
in the whole test set. HSR5 (\%) and HSR10 (\%) are also used
for SNR less than 5dB and 10dB.

\vspace{-0.1cm}
\section{Results and discussion}
\label{sec:exp}
\vspace{-0.1cm}

\subsection{Overall performance of TF-loss and DAC}
\vspace{-0.1cm}

Table \ref{tab:1} shows the performance comparison of sDPCCN with
different improvements on the whole DNS-test set, including the proposed TF-loss and
dynamic acoustic compensation, SS and MMSE-LSA. We see that the TF-loss
improves the original sDPCCN slightly due to the introduced frequency-domain
training loss. For the DAC, we investigate several setups to see its performance
behavior. DAC(J/K) means DAC with first $J$ and last $K$ frames of input noisy
speech as in Eq.(\ref{eq:tf}). DAC(UB) means the upper bound performance
of DAC, in which the ground-truth background noise of input noisy speech is directly
added into the enrollment speech to eliminate the environment mismatch.
It is obvious that the optimal configuration of DAC is DAC(4/2), the SISNR
is significantly improved from 15.12 dB to 15.96 dB, and the PESQ and DNSMOS
are also greatly improved. Moreover, when compared with DAC(UB), the proposed
simple DAC algorithm achieves very close performance to its upper bound,
which fully proves the effectiveness of our proposed DAC.

\begin{table}[!htbp]
 \vspace{-1.7em}
  \caption{Performance of sDPCCN with TF-loss and dynamic
  acoustic compensation (DAC) on DNS-test.}
  \label{tab:1}
  \centering
  \scalebox{0.97}{
  \begin{tabular}{l|c|c|c}
    \toprule
\textbf{Methods} &\;\textbf{SISNR}&\;\textbf{PESQ}&\;\textbf{DNSMOS}  \\
 & & &\;\textbf{(SIG/BAK/OVR)}  \\
    \hline
Noisy                                      &5.95  &2.16 &3.80 / 3.27 / 3.20                   \\
    \hline
sDPCCN       &15.11 &3.21 &3.79 / 3.85 / 3.32                    \\
$+$TF-loss                          &15.12 &3.28 &3.79 / 3.88 / 3.34                \\
\;\;\;\;$+$DAC(2/0)                   &15.75 &3.33 &3.79 / 3.89 / 3.35               \\
\;\;\;\;$+$DAC(2/2)                   &15.88 &3.36 &3.79 / 3.91 / 3.36            \\
\;\;\;\;$+$DAC(8/4)                 &15.81 &3.35 &3.77 / 3.91 / 3.35                  \\
\;\;\;\;$+$DAC(4/2) &\textbf{15.96} &\textbf{3.38} &\textbf{3.80 / 3.92 / 3.37}\\
\;\;\;\;$+$DAC(UB) &16.15 &3.41 &3.81 / 3.94 / 3.39\\
\hline
\;\;\;\;$+$SS(4/0)                 &15.77 &3.35 &3.76 / 3.92 / 3.35 \\
\;\;\;\;$+$MMSE-LSA                &15.01 &3.23 &3.76 / 3.92 / 3.36 \\
    \bottomrule
  \end{tabular}}
  \vspace{-10pt}
\end{table}

In addition, we also tried the spectral subtraction SS(4/0) (the best
setup) and MMSE-LSA as a pre-processing to remove the noise for comparison.
From the last two lines of Table \ref{tab:1}, we observe that
MMSE-LSA brings serious performance degradation due to the unrecoverable
speech distortion. And such distortion also makes the spectral subtraction
slightly worse than DAC(4/2). Therefore, in the following
experiments, we will directly use DAC to represent DAC(4/2)
for more concise representation.

\vspace{-0.2cm}
\subsection{Condition-wise performance of TF-loss and DAC}

\begin{table}[t]
\vspace{-1.7em}
  \caption{sDPCCN / (sDPCCN with TF-loss and DAC(4/2)) performance
  on three DNS-test subsets: t-noise, t-mix and t-nmix.}
  \label{tab:2}
  \centering
  \scalebox{0.97}{
  \begin{tabular}{c|c|c|c}
    \toprule
\textbf{Metrics} &\;\textbf{t-noise}&\;\textbf{t-mix}&\;\textbf{t-nmix}  \\
    \hline
SISNR     &16.43 / 17.36   &14.74 / 15.32   &9.26 / 10.08                   \\
    \hline
PESQ       &3.31 / 3.49  &3.32 / 3.41  &2.51 / 2.69                     \\
\hline
DNSMOS(OVR)   &3.34 / 3.38  &3.33 / 3.35  &3.22 / 3.30           \\
    \bottomrule
  \end{tabular}}
  \vspace{-12pt}
\end{table}

Table \ref{tab:2} presents the condition-wise performance comparison
between sDPCCN and the proposed sDPCCN with TF-loss and DAC 
on three subsets of DNS-test. Results show that 
both the sDPCCN baseline and proposed methods perform well on t-noise and t-mix, 
and relatively poor on t-nmix. This is due to the fact that t-nmix 
condition is more complicated than other two conditions because the 
target speech is deteriorated by both interfering speaker and noise.  
And another reason is that only a small part of nmix condition
data is included in our simulation training set.
However, when comparing the results of baseline with proposed methods, 
the performance gains on t-nmix is much larger than other conditions, 
such as, the relative (\%) SISNR/PESQ/DNSMOS improvements are 
8.9/7.2/2.5, 5.6/5.4/1.2, 3.9/2.7/0.6 
on t-nmix, t-noise and t-mix, respectively. It indicates that the introduced 
TF-loss plus DAC is more effective to improve PSE performance under 
complicated condition.

\vspace{-0.2cm}
\subsection{Examination of adaptive focal training}
\vspace{-0.1cm}

\begin{table}[htbp]
\vspace{-1.7em}
  \caption{Performance of the proposed adaptive focal training on the whole DNS-test.}
  \label{tab:3}
  \centering
  \scalebox{0.97}{
  \begin{tabular}{l|c|c|c}
    \toprule
\textbf{Methods} &\;\textbf{SISNR}&\;\textbf{PESQ}&\;\textbf{DNSMOS}  \\
 & & &\;\textbf{(SIG/BAK/OVR)}  \\
    \hline
sDPCCN       &15.11 &3.21 &3.79 / 3.85 / 3.32                    \\
$+$TF-loss                          &15.12 &3.28 &3.79 / 3.88 / 3.34                \\
\;\;\;$+$DAC &\textbf{15.96} &\textbf{3.38} &\textbf{3.80 / 3.92 / 3.37}\\
\;\;\;\;\;\;$+$AFT               &15.89 &3.36 &3.79 / 3.91 / 3.37 \\
    \bottomrule
  \end{tabular}}
  \vspace{-0.3cm}
\end{table}

Table \ref{tab:3} presents the overall performance of three proposed 
methods for improving sDPCCN. It is clear that the AFT does not provide 
any average performance gain, although it has led the model to
pay more attention to hard samples since we see a large hard sample rate 
decrease as shown in Table \ref{tab:4} (absolute 0.5, 1.25 and 
4.62 are achieved for HSR0, HSR5 and HSR10, respectively). This may 
be because the number of hard samples in the entire DNS-set is relatively small.    
In Table \ref{tab:4}, it is worth noting that TF-loss and DAC have 
also reduced the hard sample rate, but it results from the overall 
performance improvement of DNS-test, rather than especially 
improving hard samples. 

\begin{table}[htbp]
\vspace{-1.7em}
  \caption{The hard sample rate HSR0(\%), HSR5(\%) and 
  HSR10(\%) of DNS-test enhanced by different models.}
  \label{tab:4}
  \centering
  \scalebox{0.97}{
  \begin{tabular}{l|c|c|c}
    \toprule
\textbf{Methods}&\;\textbf{HSR0($\%$)}&\;\textbf{HSR5($\%$)}&\;\textbf{HSR10($\%$)}  \\
    \hline
sDPCCN            & 1.38       & 5.75      & 18.13                    \\
$+$TF-loss        & 1.50       & 5.38      & 16.63                \\
\;\;\;$+$DAC        & 1.13     & 4.00      & 13.25               \\
\;\;\;\;\;\;$+$AFT  & \textbf{0.63}   & \textbf{2.75} & \textbf{8.63}                    \\
    \bottomrule
  \end{tabular}}
  \vspace{-0.3cm}
\end{table}

\begin{figure}[t]
  \centering
  \setlength{\abovecaptionskip}{-0.2cm}
  \includegraphics[width=\linewidth]{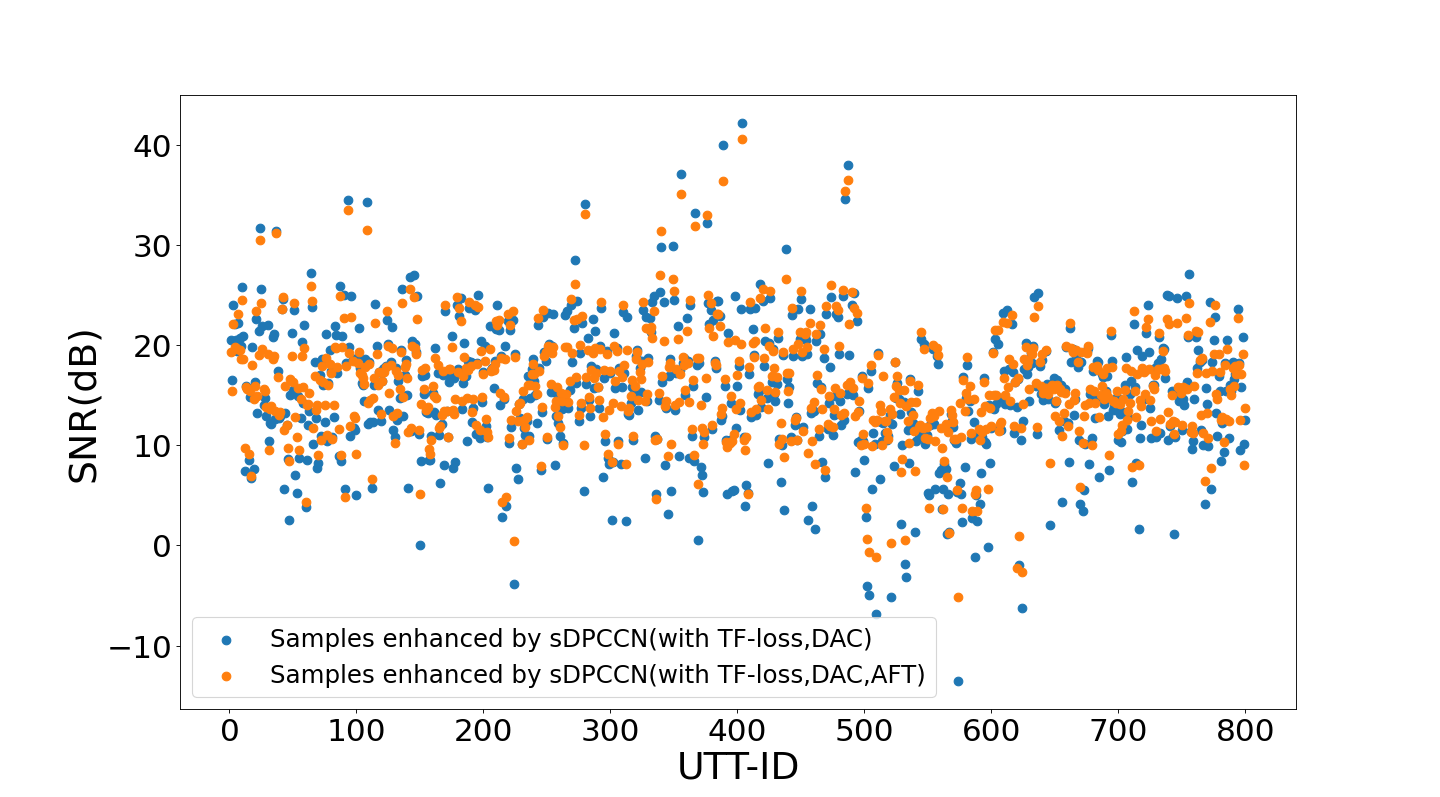}
  \caption{SNR distribution of enhanced samples in DNS-test.}
  \label{fig:3}
  \vspace{-1.3em}
\end{figure}

In Fig.\ref{fig:3}, we show the SNR distribution of all samples enhanced by 
our models (with and without AFT) in DNS-test. Compared with the blue samples (model without AFT), 
orange ones (model with AFT) are more concentrated and the number of samples 
with SNR lower than 10dB are reduced, which indicates that 
the hard samples in DNS-test are greatly improved by the proposed AFT.

\begin{figure}[htbp]
  \centering
  \includegraphics[width=7.5cm]{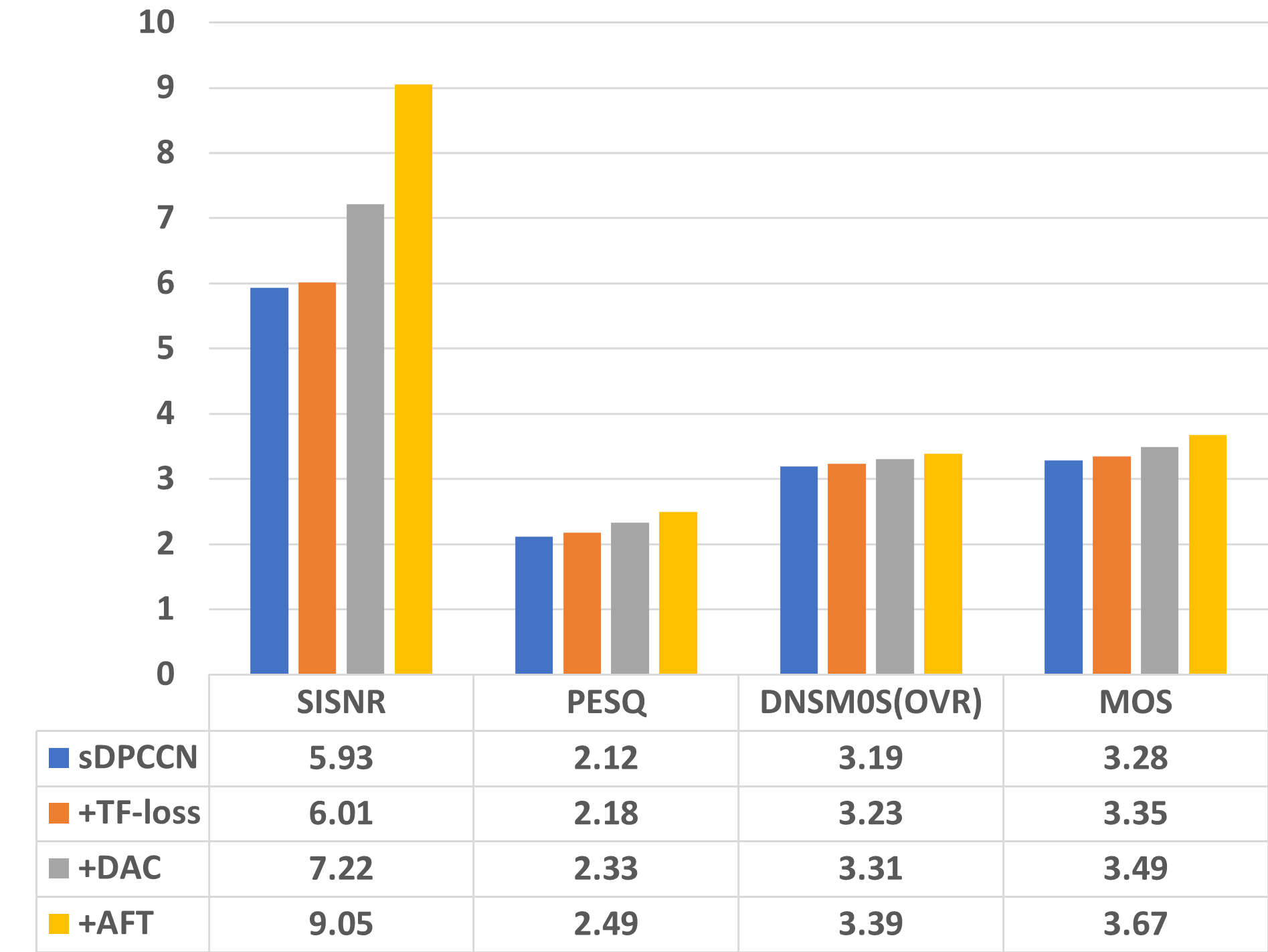}
  \caption{Performance of different models on hard sample subset.}
  \label{fig:ex}
  \vspace{-1.3em}
\end{figure}


To further prove the effectiveness of AFT for solving hard sample  
issue, in Fig.\ref{fig:ex}, we only present the performance of hard sample 
subset (samples with SNR$<$10 dB enhanced by baseline sDPCCN). Besides the SISNR, PESQ, DNSMOS, 
we also perform a listening test to give the 
traditional MOS \cite{mos} score of this small subset. Different from the 
observations in Table \ref{tab:3}, we see all the evaluation metrics are significantly 
improved by the proposed methods on this subset. In particular, we see that
the AFT brings additional absolute 1.83 dB SISNR, 0.16 PESQ, 0.08 DNSMOS and 
0.18 MOS improvements over the sDPCCN with both TF-loss and DAC. All these 
above results fully demonstrate the effectiveness of 
adaptive focal training in improving the performance of hard samples.

\vspace{-0.2cm}
\section{Conclusion}
\label{sec:conclusion}
\vspace{-0.2cm}

%

In this study, we propose three new techniques to improve our previous 
sDPCCN for personalized speech enhancement. 
Firstly, TF-loss is used to combine both time domain and frequency 
domain information during model training. Then, to alleviate the acoustic 
environment mismatch between input noise speech and clean enrollment utterance, 
a simple but effective dynamic acoustic compensation is proposed. All the 
overall and condition-wise results on the DNS-test set show that, this 
compensation brings significant improvements in terms of multiple evaluation metrics 
over original sDPCCN. Moreover, the proposed adaptive focal training is 
proved effective in different aspects for improving the hard sample performance. 
Some noisy and enhanced samples, including the DNS-test set can be found from \url{https://github.com/orcan369/DNS-test}.

\bibliographystyle{IEEEbib}
\bibliography{ICASSP2023}

\begin{thebibliography}{10}

\bibitem{mis2}
A.~Alex, L.~Wang, P.~Gastaldo, and A.~Cavallaro.
\newblock Mixup augmentation for generalizable speech separation.
\newblock In {\em International Workshop on Multimedia Signal Processing
  (MMSP)}, pages 1--6, 2021.

\bibitem{tcn}
S.~Bai, J.~Z. Kolter, and V.~Koltun.
\newblock An empirical evaluation of generic convolutional and recurrent
  networks for sequence modeling.
\newblock {\em arXiv preprint arXiv:1803.01271}, 2018.

\bibitem{6}
S.~F. Boll.
\newblock Suppression of acoustic noise in speech using spectral subtraction.
\newblock {\em IEEE/ACM Transactions on Acoustics, Speech, and Signal
  Processing}, 27:113--120, 1979.

\bibitem{model2}
L.~Chen, C.~Xu, X.~Zhang, X.~Ren, X.~Zheng, C.~Zhang, L.~Guo, and B.~Yu.
\newblock Multi-stage and multi-loss training for fullband non-personalized and
  personalized speech enhancement.
\newblock In {\em Proceedings of ICASSP}, pages 9296--9300, 2022.

\bibitem{hs3}
Y.~Cui, M.~Jia, T.~Lin, and S.~Song, Y.and~Belongie.
\newblock Class-balanced loss based on effective number of samples.
\newblock In {\em Proceedings of CVPR}, pages 9260--9269, 2019.

\bibitem{tsb}
Marc Delcroix, Tsubasa Ochiai, Katerina Zmolikova, Keisuke Kinoshita, Naohiro
  Tawara, Tomohiro Nakatani, and Shoko Araki.
\newblock Improving speaker discrimination of target speech extraction with
  time-domain speakerbeam.
\newblock In {\em Proceedings of ICASSP}, pages 691--695, 2020.

\bibitem{4}
C.~Deng, S.~Ma, Y.~Sha, Y.~Zhang, and F.~Wang.
\newblock Robust speaker extraction network based on iterative refined
  adaptation.
\newblock In {\em Proceedings of Interspeech}, pages 3530--3534, 2021.

\bibitem{hs2}
Q.~Dong, S.~Gong, and X.~Zhu.
\newblock Class rectification hard mining for imbalanced deep learning.
\newblock In {\em Proceedings of ICCV}, pages 1869--1878, 2017.

\bibitem{dns}
H.~Dubey, V.~Gopal, R.~Cutler, A.~Aazami, S.~Matusevych, S.~Braun, S.~E.
  Eskimez, M.~Thakker, T.~Yoshioka, and H.~Gamper.
\newblock {ICASSP} 2021 deep noise suppression challenge.
\newblock In {\em Proceedings of ICASSP}, pages 9271--9275, 2022.

\bibitem{lsa}
Y.~Ephraim and D.~Malah.
\newblock Speech enhancement using a minimum mean-square error log-spectral
  amplitude estimator.
\newblock 33(2):443--445, 1985.

\bibitem{data1}
J.~Gemmeke, D.~Ellis, D.~Freedman, A.~Jansen, W.~Lawrence, R.~Moore, M.~Plakal,
  and M.~Ritter.
\newblock Audio set: An ontology and human-labeled dataset for audio events.
\newblock In {\em Proceedings of ICASSP}, pages 776--780, 2017.

\bibitem{7}
J.~Han, Y.~Long, L.~Burget, and J.~Cernock.
\newblock {DPCCN}: Densely-connected pyramid complex convolutional network for
  robust speech separation and extraction.
\newblock In {\em Proceedings of ICASSP}, pages 7292--7296, 2022.

\bibitem{densenet}
G.~Huang, Z.~Liu, L.~Van Der~Maaten, and K.~Q. Weinberger.
\newblock Densely connected convolutional networks.
\newblock In {\em Proceedings of CVPR}, pages 4700--4708, 2017.

\bibitem{pesq}
I.Rec.
\newblock P.862.2: Wideband extension to recommendation p.862 for the
  assessment of wideband telephone networks and speech codecs.
\newblock {\em International Telecommunication Union,CH–Geneva}, 2005.

\bibitem{model3}
X.~Ji, M.~Yu, C.~Zhang, D.~Su, T.~Yu, X.~Liu, and D.~Yu.
\newblock Speaker-aware target speaker enhancement by jointly learning with
  speaker embedding extraction.
\newblock In {\em Proceedings of ICASSP}, pages 7294--7298, 2020.

\bibitem{tea}
Y.~Ju, W.~Rao, X.~Yan, Y.~Fu, S.~Lv, L.~Cheng, Y.~Wang, L.~Xie, and S.~Shang.
\newblock {TEA-PSE}: Tencent-ethereal-audio-lab personalized speech enhancement
  system for {ICASSP} 2022 {DNS} challenge.
\newblock In {\em Proceedings of ICASSP}, pages 9291--9295, 2022.

\bibitem{adam}
D.~Kingma and J.~Ba.
\newblock Adam: A method for stochastic optimization.
\newblock {\em Computer Science}, 2014.

\bibitem{sisnr}
J.~Le~Roux, S.~Wisdom, H.~Erdogan, and J.~R. Hershey.
\newblock {SDR}--half-baked or well done?
\newblock In {\em Proceedings of ICASSP}, pages 626--630, 2019.

\bibitem{hs1}
T.~Lin, P.~Goyal, R.~Girshick, K.~He, and P.~Dollár.
\newblock Focal loss for dense object detection.
\newblock {\em IEEE Transactions on Pattern Analysis and Machine Intelligence},
  42(2):318--327, 2020.

\bibitem{convtas}
Yi~Luo and Nima Mesgarani.
\newblock Conv-tasnet: Surpassing ideal time–frequency magnitude masking for
  speech separation.
\newblock {\em {IEEE}/{ACM} Transactions on Audio, Speech, and Language
  Processing}, 27(8):1256--1266, 2019.

\bibitem{cd3}
Y.~Lv, S.and~Hu, S.~Zhang, and Xie. L.
\newblock {DCCRN+}: Channel-wise subband dccrn with snr estimation for speech
  enhancement.
\newblock In {\em Proceedings of Interspeech}, pages 2816--2820, 2021.

\bibitem{mos}
B.~Naderi and R.~Cutler.
\newblock Subjective evaluation of noise suppression algorithms in
  crowdsourcing.
\newblock 2020.

\bibitem{mis1}
A.~Pandey and D.~Wang.
\newblock On cross-corpus generalization of deep learning based speech
  enhancement.
\newblock {\em IEEE/ACM Transactions on Audio, Speech, and Language
  Processing}, 28:2489--2499, 2020.

\bibitem{early}
L.~Prechelt.
\newblock Early stopping - but when?
\newblock {\em Springer Berlin Heidelberg}, 1998.

\bibitem{dnsmos}
C.~Reddy, V.~Gopal, and R.~Cutler.
\newblock {DNSMOS} p.835: A non-intrusive perceptual objective speech quality
  metric to evaluate noise suppressors.
\newblock In {\em Proceedings of ICASSP}, pages 886--890, 2022.

\bibitem{unet}
O.~Ronneberger, P.~Fischer, and T.~Brox.
\newblock {U-Net}: Convolutional networks for biomedical image segmentation.
\newblock In {\em Proceedings of MICCAI}, pages 234--241. Springer, 2015.

\bibitem{data2}
J~Thiemann, N.~Ito, and E.~Vincent.
\newblock The diverse environments multi-channel acoustic noise database: A
  database of multichannel environmental noise recordings.
\newblock {\em The Journal of the Acoustical Society of America}, 2013.

\bibitem{cd1}
D.~Wang, H.and~Wang.
\newblock Cross-domain speech enhancement with a neural cascade architecture.
\newblock In {\em Proceedings of ICASSP}, pages 7862--7866, 2022.

\bibitem{hsr}
K.~Wang, Y.~Peng, H.~Huang, Y.~Hu, and S.~Li.
\newblock Mining hard samples locally and globally for improved speech
  separation.
\newblock In {\em Proceedings of ICASSP}, pages 6037--6041, 2022.

\bibitem{denseunet}
Z.-Q. Wang and D.~Wang.
\newblock Count and separate: Incorporating speaker counting for continuous
  speaker separation.
\newblock In {\em Proceedings of ICASSP}, pages 11--15, 2021.

\bibitem{pyramid}
H.~Zhao, J.~Shi, X.~Qi, X.~Wang, and J.~Jia.
\newblock Pyramid scene parsing network.
\newblock In {\em Proceedings of CVPR}, pages 2881--2890, 2017.

\bibitem{cd2}
S.~Zhao, B.~Ma, K.~Watcharasupat, and W.~Gan.
\newblock {FRCRN}: Boosting feature representation using frequency recurrence
  for monaural speech enhancement.
\newblock In {\em Proceedings of ICASSP}, pages 9281--9285, 2022.

\end{thebibliography}
\end{document}